\documentclass[review]{elsarticle}
\usepackage{lineno,hyperref}
\usepackage{graphicx}
\usepackage{subcaption}
\usepackage{amsmath}
\usepackage{comment}
\usepackage{xcolor}
\usepackage{amsmath}
\usepackage{amssymb}

\begin{document}

\begin{frontmatter}

\title{Promoted current-induced spin polarization in inversion symmetry broken topological insulator thin films}


\author[1]{Maryam Heydari}

\author[1]{Hanieh Moghaddasi}

\author[1]{Mir Vahid Hosseini\corref{mycorrespondingauthor}}
\cortext[mycorrespondingauthor]{Corresponding author}
\ead{mv.hosseini@znu.ac.ir}

\author[2]{Mehdi Askari}

\address[1]{Department of Physics, Faculty of Science, University of Zanjan, P.O.Box 45196-313, Zanjan, Iran}

\address[2]{Department of Physics, Faculty of Science, Salman Farsi University of Kazerun, Kazerun 73196-73544, Iran}

\begin{abstract}
We theoretically investigate current-induced spin polarization in disordered topological insulator thin films with broken inversion symmetry under an applied in-plane electric field. Utilizing the Kubo formalism within the self-consistent Born approximation and incorporating vertex corrections to account for multiple scattering events, we analyze how disorder, chemical potential, the electrostatic potential difference between the top and bottom surfaces, and momentum-dependent hybridization affect the spin susceptibility. Our results reveal that the spin susceptibility exhibits nonzero values within a finite range around a zero gap, and this range broadens as the chemical potential increases. A higher hybridization strength induces asymmetry in the spin response. A stronger potential difference, breaking inversion symmetry, significantly enhances polarization. This enhancement is a trend attributable to band inversion and is further refined by vertex corrections. These findings provide a theoretical framework for tuning spin-charge conversion in topological thin films, with implications for spintronic device applications.
\end{abstract}

\begin{keyword}
\texttt{Spin Polarization  \sep Topological Insulator \sep Thin Film }
\end{keyword}

\end{frontmatter}

\section{Introduction}\label{sec1}

Topological insulators (TIs) have received a lot of attention for their suitable applicability in spintronic devices due to the remarkable topologically protected electronic states that are protected by time-reversal symmetry \cite{Topo1,Topo2,Topo3,Topo4,Topo5,Topo6}. TIs were initially proposed in theoretical studies, then experimentally demonstrated to exist\cite{k}. The bulk states of TIs are gapped and their Fermi level lies in the gap between the conduction and valence bands, whereas some of the surface states cross the bulk energy gap, leading to surface conduction \cite{Topobound1}. So, these materials exhibit bulk insulating properties with conductive surface states, resulting in a novel electronic phase \cite{Topobound2}. The surface states are protected by time-reversal symmetry, and as long as this symmetry is preserved, their features are robust even when imperfections are applied \cite{Symmetry}. In addition, the surface states are chiral due to a nontrivial topology originating from the strong spin-orbit interaction \cite{SpinOrbit}. Because of chirality of the surface states, the momentum and spin of the carriers are locked together. Also, the carriers of chiral surface states in TIs resemble relativistic particles in high-energy physics, so that particle-like and hole-like states touch each other at Dirac points. These features result in magneto-electric polarizability \cite{4draft}, magnetic monopole induction \cite{5draft}, and magnetic-impurity-induced local gap \cite{6draft,7draft} in TIs. These materials are now widely used in spintronic and optoelectronic devices which are based on the use of spin degree of freedom for applications of information storage and computing \cite{Topo1,Topo6}. 

When the thickness of 3D TIs becomes less than $5nm$, the topological states of opposite surfaces can hybridize leading to a gap opening in the surface spectrum \cite{thickness1} and the carriers will behave as massive Dirac fermions through the TI thin film \cite{thickness2,thickness3,thickness4}. These super thin films could be assumed to be bilayer materials that have been shown to exhibit fantastic properties \cite{imp35}. For example, conductivity and electron mobility will depend severely on thickness as soon as the thickness approaches values less than $5nm$ \cite{thickness3}. Therefore, studying the charge conductivity of magnetic TI thin films has become one of the hot topics in this field \cite{15draft}. It has been shown that TI thin films can exhibit a giant magneto-optical Kerr effect \cite{16draft}, topological magneto-electric effects \cite{17draft}, an anomalous Hall-Coulomb drag \cite{18draft}, and enhanced spin conductivity \cite{imp37}.

As already mentioned above, the main reason for the existence of TIs is the spin-orbit coupling \cite{SO}, which is a combination of spin- and orbital-freedom degrees resulting in various attractive effects \cite{b}. One of the most interesting phenomena related to spin-orbit coupling is transverse spin flows \cite{transverseSpi}. When an electric field is applied to a system in the presence of spin-orbit coupling, a transverse spin current appears, known as the spin Hall effect \cite{spinHall}. In the spin Hall effect, a net spin current is produced perpendicular to the electric field, leading to the accumulation of spins at the edges of the sample \cite{spinHall1,spinHall2}. In addition, this applied electric field can also cause the spin of the charge carriers to polarize when the carriers flow through a nonmagnetic medium \cite{f}. This phenomenon, known as the Edelstein effect \cite{Edelstein} or current-induced spin polarization (CISP) \cite{c}, arises from the asymmetric relaxation of different spin states \cite{20draft,21draft}. In practice, it can emerge through the locking of the spin to its momentum. Note that the CISP is basically different from the SHE. In the case of CISP, a homogeneous spin polarization of electrons appears throughout the sample \cite{f,e}.

CISP has been observed in various environments experimentally, including semiconductors \cite{CISPExpSemi1,CISPExpSemi2,CISPExpSemi3,CISPExpSemi4,CISPExpSemi5,CISPExpSemi6,CISPExpSemi7,CISPExp3DHole}, van der Waals heterostructures \cite{CISPExpvan1,CISPExpva2} and heavy metal thin films \cite{CISPExpHevy1,CISPExpHevy2,CISPExpHevy3}. It has been studied theoretically in a wide range of systems, ranging from two-dimensional electron gases \cite{CISPTheo2DElec1,CISPTheo2DElec2,CISPTheo2DElec3,CISPTheo2DElec4} to three-dimensional (bulk) materials \cite{CISPTheo3D1,CISPTheoQuanWell,CISPTheo3DHole1,CISPTheo3DHole2,CISPTheo3DHole3}. It has been shown that structure-property determines configurations of charge-to-spin conversion \cite{CISPTheo3DHole2}. For example, Dyrdał \textit{et al.} has investigated the CISP in graphene in the presence of the Rashba spin–orbit interaction, demonstrating that the spin polarization is perpendicular to the electric field and confined to the graphene plane \cite{dyrdal}. More recently, this work has been extended, revealing that the direction of spin polarization can be tuned by the anisotropy of the Rashba interaction \cite{hosseini}. Spin manipulation \cite{Spintronic1,Spintronic2}, modulated by electrical means, plays an essential role in the advancement of spintronics \cite{d}, particularly with the integration of novel materials such as TI \cite{SpintronicsTI} due to the high spin-orbit coupling in these materials \cite{SpinOrTI1,SpinOrTI2}. 

CISP in three-dimensional TIs has been studied both theoretically and experimentally. Recent works revealed the underling mechanisms of  CISP in TI surface states \cite{SwichCISPTI}. However, subsequent experiments demonstrated its detection on various material platforms and device architectures \cite{OriginCISPTI,DetectionCISPTI,interfaceCISPTI}. Room-temperature nonlocal spin–charge interconversion in Rashba-type systems and TI-based heterostructures was reported \cite{RoomTemCISPTI1,RoomTemCISPTI2}. Gate-controlled sign reversal of spin polarization has been realized in TI spin transistors \cite{TISpinTrans}. Several complementary detection schemes have been developed, including photoconductive differential current measurements \cite{CISPPhoton}, ferromagnetic resonance \cite{CISPReson}, transport in tunnel-coupled interfaces \cite{DetectionCISPTI,interfaceCISPTI}, and opposite spin polarizations in bulk-insulating and bulk-metallic TI flakes \cite{CISPThickOpposite}. Furthermore, a comparison of CISP has been made in the TI Bi$_2$Se$_3$, InAs Rashba states \cite{comparisonCISP}, and modulated Dirac surface states \cite{comparisonCISP0}, indicating a dominant contribution of TI surface states to CISP. Furthermore, Hwang \textit{et al.} detected electrically generated spin polarization in bulk-insulating Bi$_{1.5}$Sb$_{0.5}$Te$_{1.7}$Se$_{1.3}$ \cite{CISPBiSbTeSe}.

Burkov \textit{et al.} formulated diffusion equations for spin–charge coupled transport on TI surfaces. They showed that strong spin–momentum locking gives rise to a novel magnetoresistance effect, with the response tunable by gate voltage and relevant for spin-transistor applications \cite{CISPBurkov}. Through a linear-response density-matrix formalism, it has been shown that the Edelstein effect in TI surface states remains robust in the presence of electron–electron interactions and disorder, although both conductivity and spin polarization are renormalized \cite{CulcerCISP}. Furthermore, a comprehensive description of spin–charge interconversion in TI thin films has been provided, clarifying how structural asymmetry and surface hybridization can be exploited to improve spin–orbit torques and optimize device performance \cite{ConversionCISPTI}. In contrast, Li \textit{et al.} argued within a Boltzmann transport approach that the net ensemble spin polarization induced by current is intrinsically small, cautioning against over-interpretation of experimental signals \cite{interpretCISPTI}. Using a semiclassical treatment, it has been shown that combining impurity scattering with spin-momentum locking of the Dirac surface states, the CISP is independent of chemical potential and temperature \cite{SemiCISP}. Moreover, TI/ferromagnet heterostructures have been shown to exhibit anisotropic Edelstein effects and tunable spin–orbit torques \cite{anisotropicEdelstein}. Spin-momentum locking inhomogeneities have also been identified as a source of bilinear magnetoresistance, providing new insight into nonlinear spin–charge transport \cite{BMR}. This provides an efficient charge-to-spin conversion in TIs \cite{CISPeffic0} even in the presence of the anisotropic Fermi contours \cite{CISPeffic}.

In contrast, in ultrathin films, hybridization between top and bottom surfaces and inversion symmetry breaking introduce qualitatively new features in the spin response. Considering the features of a TI thin film, its CISP is relatively less studied \cite{CISPThinFilm}, and it really deserves further research. Here, we address this by developing a theoretical model that incorporates self-consistent impurity effects and vertex corrections. Our approach reveals how structural asymmetry and momentum-dependent coupling modulate the spin response, with implications for spintronics.


\section{Model}

\begin{figure}
    \centering
\includegraphics[width=7cm]{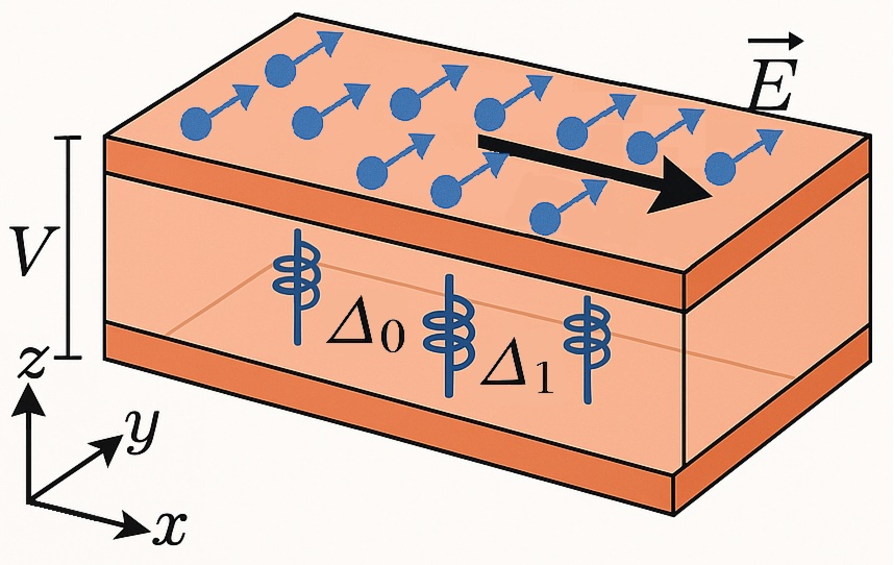}
\caption{(Color online) Schematic diagram of a TI thin film subjected to a parallel applied electric field $E$ in the $x$-direction and an electrostatic potential $V$ between the top and bottom surfaces. Also, $\Delta_{0}$ and $\Delta_{1}$ are hybridization parameters between the two surface states. The homogeneous spin polarization of electrons across the sample, resulting from the CISP effect, is depicted in blue.  
} \label{shematic}
\end{figure}

We consider a TI with thickness less than 6 quintuple layers to study CISP in the TI thin film as indicated schematically in Fig. \ref{shematic}. The TI thin film is assumed to be under an electric field, parallel to the top and bottom surfaces in a given direction such as $x$-direction. In addition, an electrostatic potential is applied perpendicularly to the surfaces in the z-direction breaking inversion symmetry. In fact, this system is similar to a bilayer in which the top and bottom surface states can hybridize. Hence, the corresponding Hamiltonian can be written as \cite{thickness3,EffecHam1,EffecHam2,highly48};
\begin{equation}
H_0(k)=v\tau_z\otimes(k_x\sigma_y-k_y\sigma_x)+\Delta(k)\tau_x\otimes\sigma_0+V\tau_z\otimes\sigma_0,
\label{Ham0}
\end{equation}
where $\sigma$ and $\tau $ are the Pauli matrices in spin and surface spaces, $k$ is in-plane momentum, and $v=\hbar\upsilon_{f}$ with $\upsilon_{f}$ being the Fermi velocity. $\Delta$ is the element of the hybridization matrix of the top and bottom surface states that is usually shown as $\Delta=\Delta_{0}-\Delta_{1}k^2$. The hybridization parameters $\Delta_{0}$ and $\Delta_{1}$ describe the tunneling between the top and bottom surfaces of the TI thin film. $\Delta_{0}$ denote the gap at zero momentum, while $\Delta_{1}$ characterizes how the gap changes with momentum. Also, $V$ is the difference in electrostatic potential between the top and bottom surfaces. The electrostatic potential difference $V$ breaks inversion symmetry, changing the states of the two surfaces. The parameters $V$, $\Delta_{0}$, and $\Delta_{1}$ determine whether the system is in a trivial or nontrivial state. When $\Delta_{0}\Delta_{1} >0$ and $V<V_c$, with $V_c=v\sqrt{|{\Delta_{0}/\Delta_{1}}|}$, being a critical potential difference, the system is topologically nontrivial, corresponds to a quantum spin Hall phase of the effective 2D system. But when the potential different increases, $V>V_c$, the system faces a transition from topological to normal insulator \cite{highly48}. 

It should be noted that the topology discussed here refers to the effective two-dimensional thin-film system formed by the hybridized top and bottom surface states, and not the original three-dimensional bulk topology of the parent topological insulator. The bulk is assumed to be a strong 3D TI, ensuring the existence of gapless surface states in the absence of hybridization. In the thin-film geometry, these states couple through tunneling, resulting in a massive Dirac Hamiltonian with the hybridization matrix term $\Delta$.  Following the analysis of \cite{highly48,thinFiTopo}, $\mathbb{Z}_2 $ invariant of this effective 2D model can be computed using the Pfaffian method.

Diagonalizing Hamiltonian (\ref{Ham0}), one gets the following eigenvalues,
\begin{equation}
  E_{lp}=l\sqrt {\Delta^2 + (v k +p V)^2}, 
     \label{BANDBAND}
\end{equation}
where $l=\pm$ indicate conduction and valence bands, and $p=\pm$ index  higher and lower bands. Throughout the paper, we take $v/a$ as the unit of energy with $a$ being the lattice constant as the unit of length.

\begin{figure}[htb]
    \centering
\includegraphics[height=7.7cm,width=7.3cm]{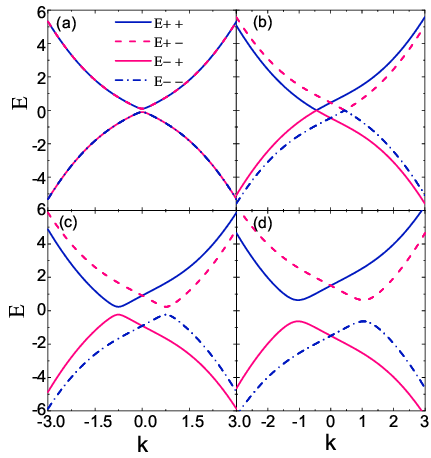}
   \caption{(Color online) The band structure of the system is shown for different values of $V$: (a) $V=0.0$ ($V<V_c$), (b) $V=0.447$ ($V=V_c$), (c) $V=0.9$ ($V>V_c$) and (d) $V=1.5$ ($V>V_c$). Here $\Delta_0=0.1$ and $\Delta_1=0.5$. }\label{fig2}
\end{figure}

The band structure of a TI thin film versus $k$ for different values of the applied potential $V$ is depicted in Fig. \ref{fig2}, illustrating how the system evolves as the potential crosses a critical threshold $V_c$. As shown in Fig. \ref{fig2}(a), at zero potential difference, $V=0$, the top and bottom surface states hybridize to open a gap at the Dirac point, preserving the inversion symmetry. Moreover, the conduction and valence bands are doubly degenerate. The resulting band structure shows a symmetric gapped dispersion around $k=0$. So, one may anticipate that the spin polarization is negligible because the contributions from both surfaces are balanced and cancel out.

At $V=V_c$, the band gap closes and the Dirac point splits into two gapless Dirac points separated horizontally with finite momentum $\pm k_D=\pm V_c/v$ giving rise to asymmetric band dispersion. The band touching points are located symmetrically around $k=0$, as shown in Fig. \ref{fig2}(b). The disappearance of the gap indicates that the hybridization and potential difference are exactly balanced, setting the stage for the transition between a topological insulator to a trivial insulator or vice versa. At this point, the band inversion is about to reverse, and the surface state symmetry is delicately balanced, causing strong sensitivity to small perturbations. Spin polarization can undergo abrupt changes here as a result of the degeneracy of spin-momentum locked states.

For $V$ values greater than $V_c$, the Dirac points go far away from each other and, at the same time, the band gap reopens but with an inverted character. [see Figs. \ref{fig2}(c) and \ref{fig2}(d)]. The conduction and valence bands that were previously associated with opposite surfaces now switch roles. This inversion introduces an asymmetry in the bands. As a result, the horizontal band splitting becomes more pronounced. Since spin-momentum locking still persists, the imbalance in contributions from top and bottom surfaces leads to enhanced net spin polarization. The spin susceptibility becomes significant in this regime, as observed in later figures.

\section{Effect of impurities}

For scattering on long-ranged impurities, we add randomly distributed disorders with density $n_{i}$ to $H_0$ giving the full Hamiltonian in the presence of disorder as
\begin{equation}
H = H_0 + V_{imp}(\mathbf{r}),
\end{equation}%
with
\begin{equation}
V_{imp}(\mathbf{r}) = V_{0}\tau_0\otimes\sigma_0\sum_{i}\delta(\mathbf{r}-\mathbf{R}_{i}),
\end{equation}%
where $V_{0}$ is the strength of the scattering potential and $\mathbf{R}_{i}$ are the scatter coordinates. We assume that the scatters are equally distributed on both surfaces and satisfy standard Born impurity correlations,
\begin{eqnarray}
\langle V_{imp}(\mathbf{r})\rangle_{c} &= & 0,\\
\langle V_{imp}(\mathbf{r})V_{imp}(\mathbf{r}')\rangle_{c} &= & n_{i}V_{0}^{2}\delta(\mathbf{r}-\mathbf{r}'),
\end{eqnarray}
where the bracket $\langle \cdots \rangle _{c}$ stands for the ensemble averaging over impurity configurations.

We also define that the impurity-averaged Green’s function which is calculated by utilization Dyson’s series that can be written as
\begin{equation}
G^\pm =\frac{G_{0}^\pm}{1-\Sigma^\pm G_{0}^\pm},\label{Dyson}
\end{equation}%
where $\Sigma^{+(-)}$ is the retarded (advanced) self‐energy and $G^{+(-)}_{0}$ is the unperturbed retarded (advanced) Green's function corresponding to the unperturbed Hamiltonian given by,
\begin{equation}
G_{0}^{\pm} = [\mu -H \pm i\eta ]^{-1}, \label{bare}
\end{equation}%
and the equation for the self-consistent self-energy is
\begin{equation}
\Sigma^\pm=\langle\tilde{V} G^{\pm} \tilde{V} \rangle =n_i V_{0}^{2}\int\frac{d^2 k}{(2\pi)^{2}} G^{\pm}. \label{selfenergy}
\end{equation}
The Green function on the right-hand side of Eq. (\ref{selfenergy}) has a matrix structure originating from the tensor product space of the layer and the spin degrees of freedom. So, the self-energy has the 4$\times$4 matrix structure as
\begin{equation}
\Sigma^\pm= \begin{bmatrix}
    \Sigma^\pm_{00} & \Sigma^\pm_{0x} & \Sigma^\pm_{0y} & \Sigma^\pm_{0z} \\
    \Sigma^\pm_{x0} & \Sigma^\pm_{xx} & \Sigma^\pm_{xy} & \Sigma^\pm_{xz} \\
    \Sigma^\pm_{y0} & \Sigma^\pm_{yx} & \Sigma^\pm_{yy} & \Sigma^\pm_{yz} \\
    \Sigma^\pm_{z0} & \Sigma^\pm_{zx} & \Sigma^\pm_{zy} & \Sigma^\pm_{zz}
  \end{bmatrix},
\end{equation}
where $\Sigma^\pm_{ij}$ with $i,j=0,x,y,z$ are the components of self-energy. Note that, in our definition, the first index refers to the layer component in $\tau$-space, and the second to the spin component in $\sigma$-space, not in the crystal axes.
Based on the above equation and according to Eq. (\ref{Dyson}), the nonzero components of the self-energy can be found as $\Sigma_{00}^{\pm}=\Sigma_{xx}^{\pm}$, $\Sigma_{yy}^{\pm}=\Sigma_{zz}^{\pm}$, and $\Sigma_{0y}^{\pm}=\Sigma_{y0}^{\pm}=\Sigma_{xz}^{\pm}=\Sigma_{zx}^{\pm}$  with
\begin{equation}
\Sigma_{0y}^{\pm} =-\frac{n_i V_{0}^{2}}{v^{2}}\int\frac{d^2 k}{(2\pi)^{2}} \frac{(\Sigma_{0y}^{\pm} + \Delta) ((\Sigma_{0y}^{\pm} + \Delta)^2 - 
\Upsilon^{\pm}\xi^{\pm} + 
   k^2 v^2)}{D^{\pm}}, \label{selfenergy0y} 
\end{equation}%
\begin{equation} 
\Sigma_{00}^{\pm} =\frac{n_i V_{0}^{2}}{v^{2}}\int\frac{d^2 k}{(2\pi)^{2}} \frac{((\Sigma_{0y}^{\pm} + \Delta)^2 -\xi^{\pm}\Upsilon^{\pm}) \Upsilon^{\pm} + 
 k^2 v^2 \xi^{\pm} }{D^{\pm}}, \label{selfenergy00} 
 \end{equation}%
\begin{equation} 
\Sigma_{zz}^{\pm} = \frac{n_i V_{0}^{2}}{v^{2}}\int\frac{d^2 k}{(2\pi)^{2}} \frac{((\Sigma_{0y}^{\pm} + \Delta)^2 - \Upsilon^{\pm}\xi^{\pm}) \xi^{\pm} + 
 k^2 v^2 \Upsilon^{\pm} }{D^{\pm}},\label{selfenergyzz} 
\end{equation}
 where 
\begin{equation}
\begin{array}{ccc}
  \Upsilon^{\pm}=(\Sigma_{zz}^{\pm} - V \mp i\eta -\mu),  &  \\
 \xi^{\pm}=(\Sigma_{00}^{\pm} + V \mp i\eta - \mu),   &  \\
 D^{\pm}=((\Sigma_{0y}^{\pm} + \Delta)^2 +(-\Upsilon^{\pm} - 
      k v) (\xi^{\pm} - 
      k v) ) ((\Sigma_{0y}^{\pm} + \Delta)^2 - (\Upsilon^{\pm} - k v) (\xi^{\pm} + 
      k v)). \label{definition}
      \end{array}
\end{equation}%
Equations (\ref{Dyson})–(\ref{definition}) describe the impurity-averaged Green’s function and self-energy corrections in detail. First, we adopt the self-consistent Born approximation, which assumes weak impurity scattering and allows us to treat disorder perturbatively; second, these equations capture both the energy renormalization and the disorder-induced broadening of the electronic states; and finally, they reveal that hybridization between the top and bottom surface states, in conjunction with impurity scattering, plays a pivotal role in modifying the band structure and consequently impacts the spin-polarization response in the TI thin film. Thus, for clarity, we decompose the self-energy as  $\Sigma^\pm=\Sigma^\prime\mp i\Gamma$ where $\Sigma^\prime$ and $\Gamma$ are the real and imaginary parts of the self-energy. Within the self-consistent treatment, we evaluated the components of the self-energy Eqs. (\ref{selfenergy0y}-\ref{selfenergyzz}).

\begin{figure}
\centering
\includegraphics[height=6cm,width=7.1cm]{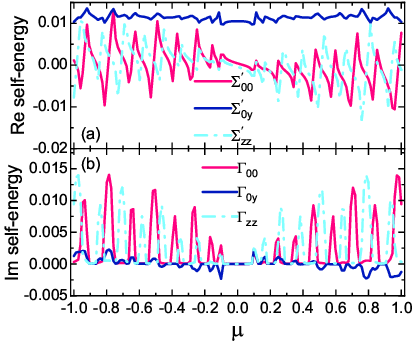}
\caption{(Color online) (a) Real and (b) imaginary parts of  $\Sigma_{00}^{-}$, $\Sigma_{0y}^{-}$ and $\Sigma_{zz}^{-}$, for $V=0.1$, $\Delta_{0}=0.1$, $\Delta_{1}=0.5$, and $n_i V_{0}^{2}=0.02$.}
\label{realselfenergy1}
\end{figure}

Figure \ref{realselfenergy1} presents the real and imaginary parts of the self-energy components $\Sigma_{00}^{-}$, $\Sigma_{0y}^{-}$ and $\Sigma_{zz}^{-}$ for a thin film of TI under inversion symmetry breaking. Panel (a) shows the real parts of self-energy terms. The real parts of the self-energy components represent the energy shifts resulting from impurity scattering and hybridization effects. Variations among $\Sigma_{00}^{'}$, $\Sigma_{0y}^{'}$ and $\Sigma_{zz}^{'}$ indicate that impurities renormalize different layer–spin channels in a nonuniform way. In particular, the difference between $\Sigma_{0y}^{'}$ and $\Sigma_{zz}^{'}$ suggests that spin channels coupled to $\sigma_y$ and $\sigma_z$ respond differently to disorder, even though the inversion-breaking potential is spin independent. Because, the potential modifies the states of the top and bottom surfaces. In the presence of spin–momentum locking, the surface-state spin texture depends on momentum and couples differently to $\sigma_x$ and $\sigma_y$ spin components. The momentum-dependent hybridization term $\Delta_1 k^2$ enhances this difference, leading to unequal self-energy components.

Panel (b) displays the imaginary parts of the self-energies. The imaginary parts of the self-energy components are directly related to the scattering rate and, hence, the finite lifetime of the electronic states. A higher magnitude in these components indicates stronger disorder-induced broadening. The distinct behavior of the different components implies that not all scattering processes equally damp the electronic states. In particular, $\Gamma_{0y}$ and $\Gamma_{zz}$ show different behaviours. It suggests that different spin–layer channels experience different levels of disorder-induced damping. In other words, the lifetime broadening is channel dependent, not uniform across all components. 

As a result, the real and imaginary parts of the self-energy components reveal channel-dependent energy shifts and scattering rates. Notably, the anisotropy in these components, particularly at finite inversion-breaking potentials, implies anisotropy in the layer$\otimes$spin basis rather than literal spatial directions. This could manifest in transport as unequal spin relaxation rates and conductivities in different spin channels, which are critical in practical TI-based spintronics systems.

\section{Current induced spin polarization}

Using the linear response theory, the electron spin polarization, $\mathbf{S}$, is related to the externally applied electric field, $\mathbf{E}$, through the relation
\begin{equation}\label{hk-matrice}
S_{\alpha}=\sum_{\beta}
 \chi_{\alpha\beta} E_{\beta},
\end{equation}
where $\chi_{\alpha\beta}$ represents the electric spin susceptibility. The spin susceptibility can be calculated using the Kubo formula yielding \cite{SpiSusc1,SpiSusc2,SpiSusc3},
\begin{equation}
\chi_{\alpha\beta} = -\frac{e\hbar }{2\pi }\int dE\frac{\partial f(E)}{%
\partial E}Tr\langle \hat{S}_{\alpha}G^{+}_{0}(E)v_{\beta}G^{-}_{0}(E)\rangle _{c},  \label{Kubo2}
\end{equation}%
where $f(E)$ is the Fermi distribution function, and $v_{\beta} = (\partial H/\partial k_{\beta})/\hbar$ is the velocity matrix along the $\beta$-th axis, with $\beta= x,y$. $Tr$ stands for trace over spin space. 

In disordered systems, impurity scattering modifies the electrons respond to external fields. In our calculation, this effect appears as a renormalization of the velocity operator, called the vertex correction. Diagrammatically, this corresponds to adding an infinite series of “ladder” diagrams, where the velocity operator is repeatedly connected by impurity lines to pairs of Green’s function. The ladder structure ensures that multiple scattering events are taken into account self-consistently. In practice, this means that the bare velocity 
$v_{\beta}$ is replaced by a dressed velocity $\tilde{v}_{\beta}$  obtained from the Bethe–Salpeter equation. Taking into account vertex corrections \cite{Edelstein,SpiSuscVertexCorr} in the ladder approximation, Eq. (\ref{Kubo2}) for $\chi_{\alpha\beta}$ at zero temperature can be recast as
\begin{equation}
\chi_{\alpha\beta} = \frac{e\hbar}{2\pi} \int \frac{d^{2}k}{(2\pi )^{2}}Tr\left[ \hat{S}
_{\alpha}G^{+}_{0}\tilde{v}_{\beta} G^{-}_{0}\right], 
\label{Kubo3}
\end{equation}
where the spin operator is defined as
\begin{equation}
\hat{S}_{\alpha} = \frac{\hbar}{2}\tau_{0}\otimes\sigma_{\alpha},
\end{equation}
with $\tau_{0}$ being a unit matrix in the layer space. In the self-consistent Born approximation, the velocity-vertex function $\tilde{v}_{\beta}$ is given by
\begin{equation} 
\tilde{v}_{\beta} =v_{\beta} + n_i V_{0}^{2}\int\frac{d^2 k}{(2\pi)^{2}} G^{+}\tilde{v}_{\beta} G^{-}.
\label{vertexcorrection1} 
\end{equation}
This formulation accounts for multiple scattering events and ensures a more accurate description of the electron dynamics in the presence of disorder.

Considering an electric field applied along the thin film plane in the x-axis direction, we calculated the corresponding $\chi$ parameters without including vertex corrections. It was found that $\chi_{xx}$ and $\chi_{zx}$ vanish upon angular averaging. While, due to symmetry constraints and spin-momentum locking, the non-zero spin susceptibility is given by $\chi_{yx}$ expressed as  
\begin{equation}
  \chi_{yx}= C
\int dk\frac{8kV \mu (V^2-\mu^2 +k^2v^2 +\Delta^2)(V^2-\mu^2 -k^2v^2 +\Delta(\Delta_{0}+3\Delta_{1}k^2))}{\prod _ {n = 1}^{4}(\mu- E_{n}+i\eta)(\mu- E_{n}-i\eta) }\notag,
\end{equation}
where $C=\frac{ev\hbar}{8\pi^{2}}$, $E_n$ are the eigenvalues of the Hamiltonian, and $\eta=\hbar/2\tau$ with $\tau$ representing the momentum-relaxation time. The integration over $k$ is carried out numerically. Thus, the spin polarization is nonzero only in the direction perpendicular to the electric field (i.e., $\chi_{yx}$). This behavior reflects the inherent anisotropy of the spin texture in topological insulators, where the spin is locked perpendicular to the momentum. In the following, we explore how $\chi_{yx}$ varies with changes in the chemical potential, the potential difference, and the hybridization parameters.

\begin{figure}
\centering
\includegraphics[height=5.2cm,width=6.7cm]{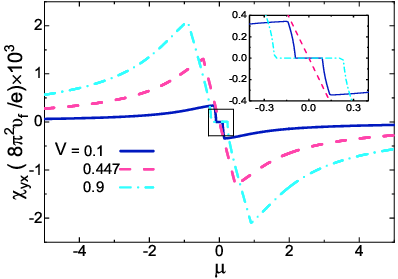}
\caption{(Color online) The spin susceptibility $\chi_{yx}$ as a function of $\mu$ for different values of potentials. Here $\Delta_0=0.1$ and $\Delta_1=0.5$. The inset shows the abrupt sign change near the Dirac point, which signals the closing of the energy gap.}
\label{mm}
\end{figure}

The spin susceptibility $\chi_{yx}$ versus the chemical potential for different values of $V$ is explored in Fig. \ref{mm}. It is evident that $\chi_{yx}$ is zero around the Dirac point, corresponding to chemical potentials within the energy gap of the band structure (see inset panel). The inset provides a close-up view around the Dirac point, where it can be seen that, at $V=V_c$ (where the band gap closes), the sign of $\chi_{yx}$ changes abruptly. For other values of the potential difference, where the gap remains open, the spin susceptibility vanishes over this range of chemical potentials. Just beyond the gap, a sudden increase in the polarization is observed, which can be attributed to the maximized energy difference between the higher (or lower) bands at these chemical potentials; in contrast, when the band separation is smaller, the upward and downward contributions to the spin polarization tend to cancel each other out. Moreover, the spin susceptibility in the conduction and valence bands exhibits equal magnitudes but opposite signs. Furthermore, an increase in the potential difference $V$ increases the magnitude of spin polarization, due to band splitting under the influence of $V$. Moreover, reversing the sign of $V$ only inverts the polarization direction, as expected (not shown).

\begin{figure}[t]
\centering
\includegraphics[height=8cm,width=10cm]{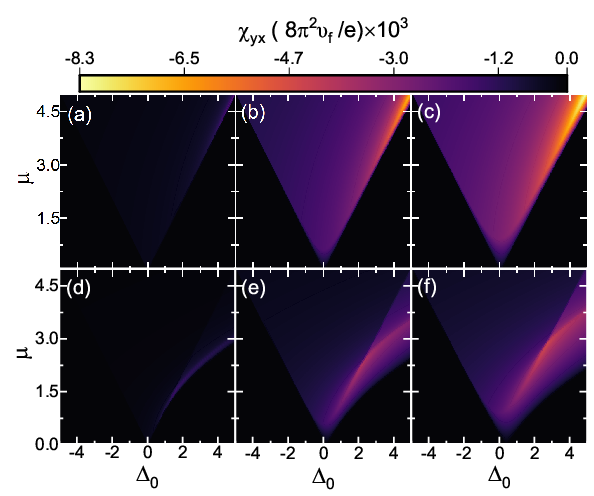}
\caption{(Color online) The spin susceptibility versus $\mu$ and $\Delta_0$ with $V=0.1$, $V=0.6$, and $V=0.9$ for the left, middle, and right columns, respectively. 
Also, $\Delta_1=0.1$ and $\Delta_1=0.5$ for the top and bottom rows, respectively.} \label{fig5}
\end{figure}

In Fig. \ref{fig5}, the spin susceptibility as functions of $\mu$ and $\Delta_0$ is depicted for different values of $V$ and $\Delta_1$. From the left column to the right column, $V$ increases, and the top and bottom rows correspond to $\Delta_1=0.1$ and $\Delta_1=0.5$, respectively. For low values of $V$ (as shown in the left column), the spin susceptibility is almost negligible and exhibits little dependence on $\mu$ and $\Delta_0$. This suggests that when the potential difference is small, the system remains nearly symmetric and the hybridized surface states yield almost canceling contributions to the overall spin polarization. In contrast, for larger values of $V$ (the middle and right columns), the potential difference induces significant band splitting in the surface state spectrum. This splitting enhances spin polarization by creating an imbalance between the contributions from different bands, thereby leading to a nonzero $\chi_{yx}$ over a finite range of $\Delta_0$. 

Moreover, as $\mu$ increases, the range of $\Delta_0$ over which the spin susceptibility is nonzero widens. This behavior can be understood by recognizing that a higher chemical potential fills more electronic states and moves the Fermi level deeper into the conduction or valence bands. In such cases, the differences in the energy dispersion (and consequently, the spin polarization) between the bands become more pronounced, allowing for a broader region where net spin polarization emerges. The transition from $\Delta_1=0.1$ (top row) to $\Delta_1=0.5$ (bottom row) leads not only to a decrease in the magnitude of spin susceptibility but also to a more asymmetric distribution of it as a function of $\Delta_0$. Physically, a larger $\Delta_1$ introduces a stronger momentum dependence in hybridization between the top and bottom surface states, effectively breaking the particle-hole symmetry more strongly. This momentum dependence means that the contributions to the spin polarization from states at different momenta become unequal, thereby shifting and weakening the overall distribution of $\chi_{yx}$ with respect to $\Delta_0$.

\begin{figure}
\centering
\includegraphics[height=8.5cm,width=10cm]{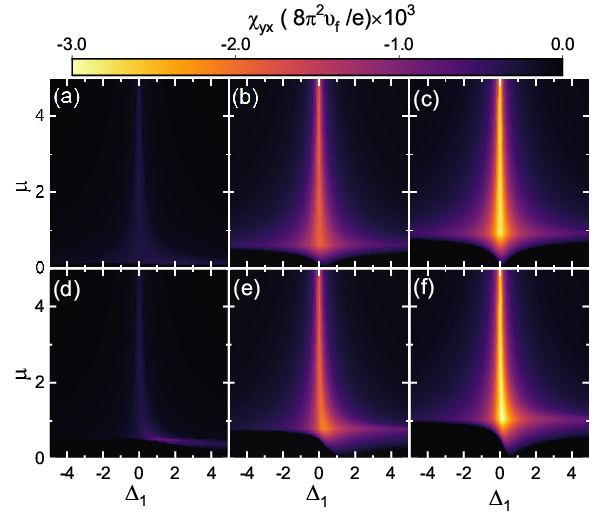}
\caption{(Color online) The spin susceptibility versus $\mu$ and $\Delta_1$ with $V=0.1$, $V=0.6$, and $V=0.9$ for the left, middle, and right columns, respectively. Also, $\Delta_0=0.1$ and $\Delta_0=0.5$ for the top and bottom rows, respectively.} \label{fig6}
\end{figure}

Figure \ref{fig6} reveals how the momentum-dependent hybridization, represented by $\Delta_1$, affects the spin susceptibility, $\chi_{yx}$, in relation to the chemical potential, $\mu$, under different applied electric potentials ($V$) and constant hybridization offset ($\Delta_0$). From the left to right columns, the potential $V$ increases, while the top and bottom rows correspond to $\Delta_0=0.1$ and $\Delta_0=0.5$, respectively. Similarly to Fig. \ref{fig5}, when $V$ is small (see the left panels), the spin susceptibility is negligible because the small potential is insufficient to induce significant band splitting. As the applied potential increases (middle and right columns), the band structure is more strongly modulated, leading to a pronounced imbalance in the population of the spin-polarized states. This enhanced band splitting yields higher values of $\chi_{yx}$, clearly reflecting the role of the applied potential in generating net spin polarization.

When comparing the top row ($\Delta_0=0.1$) with the bottom row ($\Delta_0=0.5$), one observes that a larger $\Delta_0$ tends to shift the energy levels of the hybridized surface states near the band edge. This shift affects the balance between spin-up and spin-down contributions and leads to a more pronounced asymmetric spin susceptibility profile with respect to $\Delta_1$ especially at lower values of $\mu$. Additionally, for certain values of $\mu$ around the band edge, the spin susceptibility attains moderate values at finite $\Delta_1$. Because as $\mu$ increases, the Fermi level moves into regions where the energy dispersion of the hybridized bands changes more dramatically with $\Delta_1$. Thus, the differences in the contributions from various momentum states grow, further enhancing the net $\chi_{yx}$.     

\begin{figure}[htb]
\centering
\includegraphics[height=8cm,width=10cm]{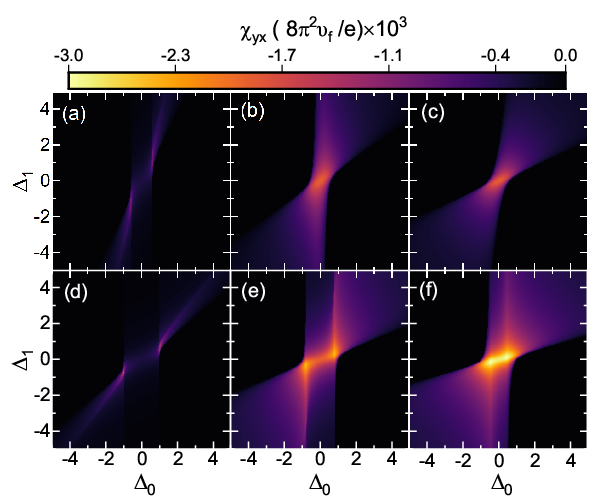}
\caption{(Color online) The spin susceptibility versus $\Delta_0$ and $\Delta_1$ with $V=0.1$, $V=0.6$, and $V=0.9$ for the left, middle, and right columns, respectively. Also, $\mu=0.6$ and $\mu=1$ for the top and bottom rows, respectively.}
\label{fig7}
\end{figure}

Finally, the spin susceptibility behavior as a function of both hybridization parameters, $\Delta_0$ and $\Delta_1$ is illustrated in Fig. \ref{fig7} for several values of the applied potential $V$ and two representative chemical potentials. In Figs. \ref{fig7}(a)–\ref{fig7}(c), the chemical potential is fixed at $\mu=0.6$ while the potential $V$ is varied (0.1, 0.6, and 0.9, respectively). In Fig. \ref{fig7}(a), two symmetric peaks in spin susceptibility are clearly observed. These peaks indicate that the contributions from the two surfaces are balanced. As $V$ increases, the band inversion becomes more pronounced. This enhances the spin susceptibility peaks. In particular, when $V$ approaches $\mu$, the two peaks draw closer until they merge into a single broader peak, a signature of a critical rearrangement of the band structure and the associated spin polarization. 

In the lower set of panels, Figs. \ref{fig7}(d)–\ref{fig7}(f), where $\mu$ is set to 1, the Fermi level lies deeper into the conduction (or valence) band, and $V$ is again varied as 0.1, 0.6, and 0.9, the two peaks become not only more pronounced but also more widely separated, as seen in Fig. \ref{fig7}(d). This indicates that the effective imbalance between the contributions from the upper and lower bands is accentuated at higher chemical potentials. Essentially, the deeper Fermi level enhances the differentiation of spin-resolved states, leading to a more sensitive dependence of the spin response on the hybridization parameters. With further an increase in $V$, the peaks gradually approach each other around the origin, as depicted in Figs. \ref{fig7}(e) and \ref{fig7}(f).

\section{Vertex corrections}

In the presence of vertex corrections, using Eq. (\ref{vertexcorrection1}) and iterating, the first-order single-impurity velocity-vertex function $\tilde{v}_{\beta}^{(1)}$ can be derived as \cite{(Japanese article)}
\begin{equation}
\tilde{v}_{\beta}^{(1)} = n_i V_{0}^{2}\int\frac{d^2 k}{(2\pi)^{2}} G^{+} v_{\beta} G^{-}. \label{firstordervelocity} 
\end{equation}
After substituting the retarded and advanced Green’s functions into the above equation, the dominant terms simplify to 
\begin{equation}
\begin{array}{cc}
\tilde{v}_{x}^{(1)} = -\delta v_{f}\tau_z\otimes\sigma_y, &  \\
\tilde{v}_{y}^{(1)} = +\delta v_{f}\tau_z\otimes\sigma_x, \label{vertexcorrection} 
      \end{array}
\end{equation}%
where $\delta$ is the velocity correction. Consequently, in the weak scattering limit, the velocity vertex can be expanded as 
\begin{equation}
\begin{array}{cc}
\tilde{v}_{x} = v_{x}-\delta v_{f}\tau_z\otimes\sigma_y, &  \\
\tilde{v}_{y} = v_{y}+\delta v_{f}\tau_z\otimes\sigma_x. \label{vertexcorrection2} 
      \end{array}
\end{equation}
By substituting Eqs. (\ref{vertexcorrection2}) into Eq. (\ref{vertexcorrection1}), the velocity correction $\delta$ can be computed numerically (see the Appendix).

\begin{figure}[htb]
\centering
\includegraphics[height=5.18cm,width=10cm]{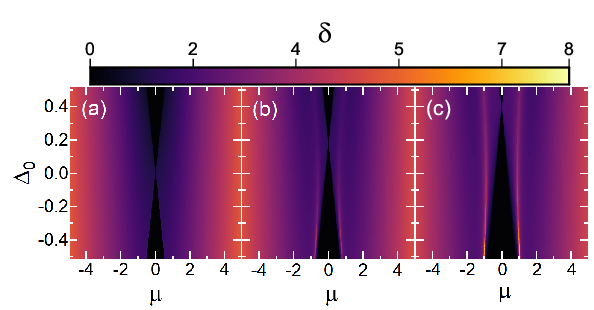}
\caption{(Color online) The vertex correction versus $\Delta_0$ and $\mu$ for $\Delta_1=0.5$ in (a) $V=0.1$, (b) $V=0.6$, and (c) $V=0.9$. Here $n_{i}V_{0}^2=0.02$.}
\label{delta-mu,d0-d1=0.5-separate1}
\end{figure}

Figure \ref{delta-mu,d0-d1=0.5-separate1} illustrates the variation of the vertex correction $\delta$ as functions of $\Delta_0$ and $\mu$ for different values of the potential $V$: (a) $V = 0.1$, (b) $V = 0.6$, and (c) $V = 0.9 $. Here, the hybridization parameter is set to $\Delta_1 = 0.5$. At low values of $V$, the system remains almost symmetric. In this regime, the vertex correction $\delta$ is small or nearly uniform across the ($\Delta_0$, $\mu$) plane, except for the gapped region where $\delta$ is zero. The weak band splitting implies that impurity scattering contributes symmetrically from both surfaces, resulting in minimal velocity correction. Therefore, impurity-induced modifications to the current response are mild and smoothly varying. As $V$ increases to intermediate and high values (0.6 and 0.9, respectively), the band structure becomes increasingly asymmetric due to stronger electrostatic splitting, leading to enhanced and nonuniform scattering processes. This causes $\delta$ to grow significantly and exhibit sharp features, particularly near the band edges and gap-closing regions. Notably, the region of finite $\delta$ shifts toward larger $\Delta_0=0$ with increasing $V$. This shift can be understood as a consequence of the competition between the hybridization-induced gap ($\Delta_0$) and the potential difference $V$. Consequently, as the band structure becomes more inverted (owing to the large $V$), the system's response to disorder becomes more sensitive to $\Delta_0$  and $\mu$.

\begin{figure}[htb]
\centering
\includegraphics[height=5.2cm,width=10cm]{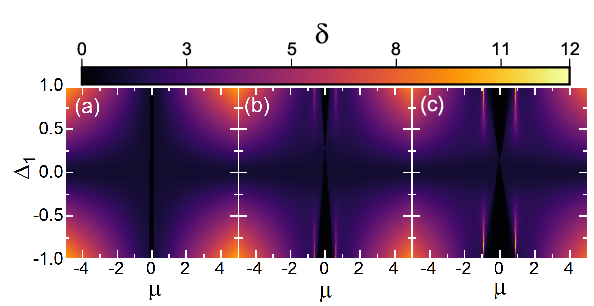}
\caption{(Color online) The vertex correction $\delta$ versus $\Delta_1$ and $\mu$ for $\Delta_0=0.1$ in (a) $V=0.1$, (b) $V=0.6$, and (c) $V=0.9$. Here $n_{i}V_{0}^2=0.02$.}
\label{delta-mud1-d0=0.1}
\end{figure}

Figure \ref{delta-mud1-d0=0.1} shows the vertex correction $\delta$ as functions of $\Delta_1$ and $\mu$ for a fixed $\Delta_0=0.1$ and three different values of the potential $V$: (a) $V = 0.1$, (b) $V = 0.6$, and (c) $V = 0.9$. The hybridization parameter $\Delta_0$ is set at $0.1$. For low values of $\Delta_1$, the vertex correction is small and almost independent of $\mu$. However, as $\Delta_1$ increases, the vertex correction increases accordingly away from the charge neutrality point. This behavior can be understood as follows. The momentum-dependent hybridization term $\Delta_1$ introduces an additional $k^2$ correction to the hybridization gap, so that a higher $\Delta_1$ enhances the differences between contributions from various momentum states. This induces a stronger asymmetry in the band structure and spin texture at higher energies. This amplifies the effect of impurity scattering at finite chemical potentials, making vertex corrections more significant in the doped regime. However, near the charge neutrality point, the relevant $k$-states are small, so the momentum-dependent part of the hybridization ($\Delta_1 k^2$) contributes little.
Thus, the vertex correction remains relatively small near $\mu=0$, even for large $\Delta_1$, since the bands retain more symmetry there and the system is less sensitive to $\Delta_1$.

Moreover, there is a gap region around $\mu=0$, due to $\Delta_0$, where the vertex correction is vanishingly small. At low values of $V$ (e.g., $V=0.1$), around the band edge, the vertex correction $\delta$ is small and almost independent of $\mu$. This suggests that when the potential difference is weak, impurity scattering affects all states fairly uniformly. However, as the applied potential $V$ increases (moving to $V=0.6$ and $V=0.9$), this gap region expands, due to the increasing gap, particularly at the grater absolute values of $\Delta_1$. At the same time, a considerable vertex correction appears at the boundaries of this region. Because, at such potentials the band structure becomes more asymmetric and inverted. Subsequently, the density of states and the associated spin textures change more rapidly. Consequently, $\delta$ shows a noticeable dependence on $\mu$, reflecting that impurity scattering—and hence the renormalized velocity—varies significantly as the Fermi level moves within the bands.

\begin{figure}[htb]
\centering
\includegraphics[height=5.44cm,width=10cm]{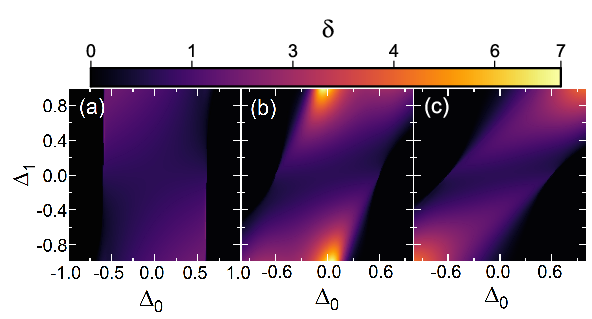}
\caption{(Color online) The vertex correction $\delta$ is investigated versus $\Delta_0$ and $\Delta_1$ simultaneously for (a) $V=0.1$, (b)$V=0.6$, and (c) $V=0.9$. Here $\mu=0.6$ and $n_{i}V_{0}^2=0.02$.  }\label{delta-d0d1times0.02}
\end{figure}

Figure \ref{delta-d0d1times0.02} presents the variation of the vertex correction $\delta$ as functions of both $\Delta_0$ and $\Delta_1$ for different values of the potential $V$: (a) $V = 0.1$, (b) $V = 0.6$, and (c) $V = 0.9$. The chemical potential is fixed at $\mu = 0.6$. For low potential ($V = 0.1$), the overall band structure remains relatively symmetric, and the vertex correction $\delta$ increases only gradually with $\Delta_0$ and $\Delta_1$. However, at higher potentials ($V = 0.6$ and $V = 0.9$), the increase in $\delta$ is much more significant, particularly in regions where both hybridization parameters are large in magnitude. This can be attributed to the fact that the external electric field drives significant band inversion and enhances the intrinsic asymmetry of the system. Under these conditions, the scattering processes become more anisotropic because the energy separation and momentum-dependence are more pronounced. Therefore, the combined effects of $\Delta_0$, $\Delta_1$, and $V$ play a crucial role in the renormalization of the electron velocity by impurity, suggesting a strong dependence of the spin response on the hybridization strength and applied potential.

\section{Conclusions}\label{sec4}

We have developed a comprehensive theoretical framework to investigate CISP in topological insulator thin films with broken inversion symmetry. Using the Kubo formalism within the self-consistent Born approximation and including vertex corrections, we analyzed how the transverse spin susceptibility $\chi_{yx}$ is influenced by key factors, such as chemical potential ($\mu$), potential difference ($V$), hybridization parameters ($\Delta_0$ and $\Delta_1$), and disorder. 

Our results show that a finite spin susceptibility emerges in a well-defined chemical potential window near the charge neutrality point, and this window broadens with increasing $\mu$. The inclusion of a momentum-dependent hybridization term ($\Delta_1$) induces pronounced asymmetry in $\chi_{yx}$. As the inversion-breaking potential $V$ increases, band inversion becomes stronger, amplifying the magnitude of spin polarization and shifting the peak features of $\chi_{yx}$. We also found that vertex corrections significantly enhance the spin response, particularly in regimes with large hybridization and strong inversion asymmetry.

Although our study is based on a generic model, for the relevant scales of the parameters, we refer to recent first-principles modeling of Bi$_2$Se$_3$, Bi$_2$Te$_3$, and Sb$_2$Te$_3$ thin
films \cite{Zsurka2024,highly48}. For Bi$_2$Se$_3$ thin films with
thicknesses 3--6 QLs, the fitted momentum-independent gap takes values $\Delta_0 \approx 3$--$120$ meV (corresponding to $\Delta_0 \approx 0.006$--$0.10$ in dimensionless units), while the momentum-dependent gaps are $\Delta_1 \approx -16$ eV\AA$^2$ (corresponding to $\Delta_1 \approx -2.0$). For Bi$_2$Te$_3$, the hybridization gap is smaller, $\Delta_0 \approx 1.5$--$14$ meV ($\Delta_0 \approx 0.005$--$0.07$), with $\Delta_1 \approx -28$ to $-30$ eV\AA$^2$ ($\Delta_1 \approx -5$ to $-7$). Sb$_2$Te$_3$ films show somewhat larger gaps at small thickness, $\Delta_0 \approx 6$--$32$ meV
($\Delta_0 \approx 0.005$--$0.05$), and $\Delta_1 \approx -13$ to $-16$ eV\AA$^2$ ($\Delta_1 \approx -1.0$ to $-1.3$). Remarkably, for ultrathin films with a thickness of 2 QLs, in the cases of Bi$_2$Se$_3$ and Sb$_2$Te$_3$, the above-mentioned values can be up to one order of magnitude larger \cite{Zsurka2024}. Thus, the parameter windows we employ map directly onto the experimentally accessible regimes.

Finally, our predicted features of current-induced spin polarization can be directly probed by current experimental statues such as spin potentiometry \cite{LiNatNano2014} and spin-torque ferromagnetic resonance \cite{MellnikNature2014}. These methods have already been applied to Bi$_2$Se$_3$ and Bi$_2$Te$_3$ films. Circularly polarized photoconductivity \cite{CISPPhoton} and nonlocal spin–charge conversion setups \cite{OriginCISPTI} provide additional approaches to probe the role of inversion asymmetry and thickness.

\section*{Acknowledgement}
We are grateful to T. Chiba and A. Shitade for valuable discussions.

\appendix

\appendix
\section*{Appendix: Derivation of the velocity vertex corrections}

In the ladder approximation, the dressed velocity $\tilde v_x$ is determined by
\begin{equation}
\tilde v_x = v_x + n_i V_0^2 \int\!\frac{d^2k}{(2\pi)^2}\, G^{+}\, \tilde v_x \, G^{-}.
\label{eq:ladder-appendix}
\end{equation}
We expand the dressed vertex in the 16-matrix basis $\{\tau_i\!\otimes\!\sigma_j\}$. The in-plane rotational symmetry, time-reversal symmetry, and scalar disorder restrict the allowed channels. For the present case only two basis elements mix under the ladder kernel:

\begin{equation}
\tilde{v}_x
= a\,(\tau_z \otimes \sigma_y)
+ b\,(\tau_z \otimes \sigma_x)
- \frac{2 k_x \Delta_1}{\hbar}\,(\tau_x \otimes \sigma_0).
\end{equation}
Projecting Eq.~\eqref{eq:ladder-appendix} onto these two channels leads to the $2\times2$ linear system
\begin{equation}
\begin{pmatrix} a \\ b \end{pmatrix}
=
\begin{pmatrix} v \\ 0 \end{pmatrix}
+
\begin{pmatrix}
\mathcal I_{yy} & \mathcal I_{yx} \\
\mathcal I_{xy} & \mathcal I_{xx}
\end{pmatrix}
\begin{pmatrix} a \\ b \end{pmatrix},
\label{eq:matrix-system}
\end{equation}
where the kernels $\mathcal I_{\alpha\beta}$ are defined by
\begin{equation}
\mathcal I_{\alpha\beta}
= n_i V_0^2 \int \frac{d^2k}{(2\pi)^2}
\, \mathrm{Tr}\!\left[ \mathcal P_\alpha \, G^{+}\, \mathcal P_\beta \, G^{-} \right].
\label{eq:integrals}
\end{equation}
Here $\mathcal P_\alpha$ are projection operators onto the relevant basis elements:
\begin{align}
\mathcal P_{yy} &= \tfrac{1}{4} (\tau_z \otimes \sigma_y), \qquad
\mathcal P_{xx} = \tfrac{1}{4} (\tau_z \otimes \sigma_x), \\
\mathcal P_{yx} &= \tfrac{1}{4} (\tau_z \otimes \sigma_x), \qquad
\mathcal P_{xy} = \tfrac{1}{4} (\tau_z \otimes \sigma_y).
\end{align}
Because the system is rotationally invariant in the $x$--$y$ plane, the integrals reduce to
\begin{equation}
\mathcal I_{\alpha\beta} = \frac{n_i V_0^2}{2\pi} \int_0^\infty \! \frac{k\, dk}{(2\pi)} \,
\langle \mathrm{Tr}[ \mathcal P_\alpha G^+(\mu,k,\theta)\, \mathcal P_\beta G^-(\mu,k,\theta) ] \rangle_\theta ,
\end{equation}
where $\langle \cdots \rangle_\theta$ denotes averaging over the polar angle $\theta$ of $\mathbf k$. 

Using the spectral representation of the Green's function,
\begin{equation}
G^{\pm} = \sum_{n=1}^4 \frac{ |u_{n,\mathbf k}\rangle \langle u_{n,\mathbf k}| }{\mu - E_n(k) \pm i\eta},
\end{equation}
with eigenvalues
\begin{equation}
E_{l,p}(k) = l \sqrt{ \Delta(k)^2 + (v k + p V)^2}, \quad l,p = \pm 1,
\end{equation}
and eigenvectors $|u_{n,\mathbf k}\rangle$
the kernels become
\begin{equation}
\mathcal I_{\alpha\beta} = \frac{n_i V_0^2}{(2\pi)^2} \sum_{n,m}\int_0^\infty \! k\, dk \,
\frac{\langle u_{n,\mathbf k}|\mathcal P_\alpha|u_{m,\mathbf k}\rangle \, 
\langle u_{m,\mathbf k}|\mathcal P_\beta|u_{n,\mathbf k}\rangle}
{(\mu - E_n(k) + i\eta)(\mu - E_m(k) - i\eta)} .
\label{eq:spectral-integral}
\end{equation}

For numerical purposes, the angular average reduces most cross terms to zero by symmetry, leaving only diagonal contributions in $\sigma_y$ and $\sigma_x$ channels. The $k$-integrals are convergent and can be evaluated numerically. In practice, $\mathcal I_{xy}\simeq\mathcal I_{yx}\simeq 0$, so the system \eqref{eq:matrix-system} reduces approximately to $a=v_f/(1-\mathcal I_{yy})$ and $b\simeq 0$, with $a=v_f/(1+\delta)
$.



\begin{thebibliography}{99} 

\bibitem{Topo1}
M.Z. Hasan and C.L. Kane, Colloquium: topological insulators, Rev. Mod. Phys. 82 (4) (2010) 3045.

\bibitem{Topo2}
M.Z. Hasan and J.E. Moore, Three-dimensional topological insulators, Annu. Rev. Condens. Matter Phys. 2 (1) (2011) 55-78.

\bibitem{Topo3}
M. Leijnse and K. Flensberg, Introduction to topological superconductivity and Majorana fermions, Semicond. Sci. Technol. 27 (12) (2012) 124003.

\bibitem{Topo4} 
X.-L. Qi and S.-C. Zhang, Topological insulators and superconductors, Rev. Mod. Phys. 83 (2011) 1057-1110.

\bibitem{Topo5} 
Y. Ando, Topological insulator materials, J. Phys. Soc. Jpn. 82 (10) (2013) 102001.

\bibitem{Topo6}
C.-K. Chiu, J.C.Y. Teo, A.P. Schnyder, and S. Ryu, Classification of topological quantum matter with symmetries, Rev. Mod. Phys. 88 (3) (2016) 035005.

\bibitem{k}
D. Hsieh, D. Qian, L. Wray, Y. Xia, Y.S. Hor, R.J. Cava, and M.Z. Hasan, A topological Dirac insulator in a quantum spin Hall phase, Nature 452 (7190) (2008) 970-974.

\bibitem{Topobound1}
J.E. Moore, The birth of topological insulators, Nature 464 (7286) (2010) 194-198. 

\bibitem{Topobound2}
X.-L. Qi and S.-C. Zhang, The quantum spin Hall effect and topological insulators, Phys. Today 63 (1) (2010) 33-38.

\bibitem{Symmetry}
A.P. Schnyder, S. Ryu, A. Furusaki, and A.W.W. Ludwig, Classification of topological insulators and superconductors in three spatial dimensions, Phys. Rev. B 78 (19) (2008) 195125.

\bibitem{SpinOrbit}
A. Manchon, H.C. Koo, J. Nitta, S.M. Frolov, and R.A. Duine, New perspectives for Rashba spin-orbit coupling, Nat. Mater. 14 (9) (2015) 871-82.

\bibitem{4draft}
A.M. Essin, J.E. Moore, and D. Vanderbilt, Magnetoelectric polarizability and axion electrodynamics in crystalline insulators, Phys. Rev. Lett. 102 (14) (2009) 146805.

\bibitem{5draft}
X.-L. Qi, R. Li, J. Zang, and S.-C. Zhang, Inducing a magnetic monopole with topological surface states, Science 323 (5918) (2009) 1184-1187.

\bibitem{6draft}
Q. Liu, C.-X. Liu, C. Xu, X.-L. Qi, and S.-C. Zhang, Magnetic impurities on the surface of a topological insulator, Phys. Rev. Lett. 102 (15) (2009) 156603.

\bibitem{7draft}
D.A. Abanin and D.A. Pesin, Ordering of magnetic impurities and tunable electronic properties of topological insulators, Phys. Rev. Lett. 106 (13) (2011) 136802.

\bibitem{thickness1}
G. Zhang, H. Qin, J. Teng, J. Guo, Q. Guo, X. Dai, Z. Fang, and K. Wu, Quintuple-layer epitaxy of high-quality Bi$_2$Se$_3$ thin films of topological insulator, Appl. Phys. Lett. 95 (5) (2009) 053114.

\bibitem{thickness2}
H.-Z. Lu, W.-Y. Shan, W. Yao, Q. Niu, and S.-Q. Shen, Massive Dirac fermions and spin physics in an ultrathin film of topological insulator, Phys. Rev. B 81 (11) (2010) 115407.

\bibitem{thickness3}
Y. Zhang, K. He, C.-Z. Chang, C.-L. Song, L. Wang, X. Chen, J. Jia, Z. Fang, X. Dai, W.-Y. Shan, and et al., Crossover of the three-dimensional topological insulator Bi$_2$Se$_3$ to the two-dimensional limit, Nat. Phys. 6 (2010) 584–588.

\bibitem{thickness4}
D. Kim, P. Syers, N.P. Butch, J. Paglione, and M.S. Fuhrer, Coherent topological transport on the surface of Bi$_2$Se$_3$, Nat. Commun. 4 (1) (2013) 2040.

\bibitem{imp35}
S.S. Pershoguba and V.M. Yakovenko, Spin-polarized tunneling current through a thin film of a topological insulator in a parallel magnetic field, Phys. Rev. B 86 (16) (2012).

\bibitem{15draft}
C.-Z. Chang, J. Zhang, X. Feng, J. Shen, Z. Zhang, M. Guo, K. Li, Y. Ou, P. Wei, L.-L. Wang, and et al., Experimental observation of the quantum anomalous Hall effect in a magnetic topological insulator, Science 340 (6129) (2013) 167-170.

\bibitem{16draft}
W.-K. Tse and A.H. MacDonald, Giant magneto-optical Kerr effect and universal Faraday effect in thin-film topological insulators, Phys. Rev. Lett. 105 (5) (2010) 057401.

\bibitem{17draft}
T. Morimoto, A. Furusaki, and N. Nagaosa, Topological magnetoelectric effects in thin films of topological insulators, Phys. Rev. B 92 (8) (2015) 085113.

\bibitem{18draft}
H. Liu, W.E. Liu, and D. Culcer, Anomalous Hall Coulomb drag of massive Dirac fermions, Phys. Rev. B 95 (20) (2017) 205435.

\bibitem{imp37}
S. Pooyan and M.V. Hosseini, Enhanced and stable spin Hall conductivity in a disordered time-reversal and inversion symmetry broken topological insulator thin film, Sci. Rep. 12 (1) (2022) 15379.

\bibitem{SO}
R. Winkler, Spin-orbit coupling effects in two-dimensional electron and hole systems, Springer-Verlag, Berlin, 2003.

\bibitem{b} 
M.I. Dyakonov and V.I. Perel, Current-induced spin orientation of electrons in semiconductors,  Phys. Lett. A 35 (6) (1971) 459-460.

\bibitem{transverseSpi} 
M.I. D’yakonov and V.I. Perel’, Possibility of orienting electron spins with current, JETP Lett. 13 (1971) 467.

\bibitem{spinHall}
J. Sinova, S.O. Valenzuela, J. Wunderlich, C. Back, and T. Jungwirth, Spin hall effects, Rev. Mod. Phys. 87 (4) (2015) 1213-1260.

\bibitem{spinHall1}
J.E. Hirsch, Spin hall effect, Phys. Rev. Lett. 83 (1999) 1834–1837.

\bibitem{spinHall2}
Y.K. Kato, R.C. Myers, A.C. Gossard, and D.D. Awschalom, Observation of the Spin Hall Effect in Semiconductors, Science 306 (5703) (2004) 1910–1913.

\bibitem{f}
E.L. Ivchenko and G.E. Pikus, New photogalvanic effect in gyrotropic crystals, JETP Lett. 27 (11) (1978) 604-608.

\bibitem{Edelstein}
V.M. Edelstein, Spin polarization of conduction electrons induced by electric current in two-dimensional asymmetric electron systems, Solid State Commun. 73 (1990) 233-235.

\bibitem{c}
S.D. Ganichev, M. Trushin, and J. Schliemann, Spin polarisation by current, arXiv:1606.02043 (2016).

\bibitem{20draft}
N.S. Averkiev, L.E. Golub, and M. Willander, Spin relaxation anisotropy in two-dimensional semiconductor systems, J. Phys.: Condens. Matter. 14 (12) (2002) R271.

\bibitem{21draft}
V.L. Korenev, Bulk electron spin polarization generated by the spin Hall current, Phys. Rev. B 74 (4) (2006) 041308.

\bibitem{e}
A.G. Aronov, Y.B. Lyanda-Geller, G.E. Pikus, and D. Parsons, Spin polarization of electrons by an electric current, Sov. Phys. JETP 73 (1991) 537-541.

\bibitem{CISPExpSemi1}
Y.K. Kato, R.C. Myers, A.C. Gossard, and D.D. Awschalom, Current-Induced Spin Polarization in Strained Semiconductors, Phys. Rev. Lett. 93 (2004) 176601.

\bibitem{CISPExpSemi2}
A.Y. Silov, P.A. Blajnov, J.H. Wolter, R. Hey, K.H. Ploog, and N.S. Averkiev, Current-induced spin polarization at a single heterojunction, Appl. Phys. Lett. 85 (24) (2004) 5929-5931.

\bibitem{CISPExpSemi3}
V. Sih, R.C. Myers, Y.K. Kato, W.H. Lau, A.C. Gossard, and D.D. Awschalom, Spatial imaging of the spin Hall effect and current-induced polarization in two-dimensional electron gases, Nat. Phys. 1 (2005) 31–35.

\bibitem{CISPExpSemi4}
S.D. Ganichev, S.N. Danilov, P. Schneider, V.V. Bel’kov,
L.E. Golub, W. Wegscheider, D. Weiss, and W. Prettl, Electric current-induced spin orientation in quantum well structures, Journal of Magnetism and Magnetic Materials 300 (2006) 127-131.

\bibitem{CISPExpSemi5}
C.L. Yang, H.T. He, L. Ding, L.J. Cui, Y.P. Zeng, J.N. Wang, and W.K. Ge, Spectral Dependence of Spin Photocurrent and Current-Induced Spin Polarization in an InGaAs/InAlAs Two-Dimensional Electron Gas, Phys. Rev. Lett.
96 (2006) 186605.

\bibitem{CISPExpSemi6}
H.J. Chang, T.W. Chen, J.W. Chen, W.C. Hong,
W.C. Tsai, Y.F. Chen, and G.Y. Guo, Current and Strain-Induced Spin Polarization in InGaN/GaN Superlattices, Phys. Rev. Lett.
98 (2007) 136403.

\bibitem{CISPExpSemi7}
B.M. Norman, C.J. Trowbridge, D.D. Awschalom,
and V. Sih, Erratum: Current-Induced Spin Polarization in Anisotropic Spin-Orbit Fields, Phys. Rev. Lett. 112 (5) (2014) 056601.

\bibitem{CISPExp3DHole}
T. Furukawa, Y. Shimokawa, K. Kobayashi, and T. Itou, Observation of current-induced bulk magnetization in elemental tellurium, Nat. Commun. 8 (1) (2017) 954.

\bibitem{CISPExpvan1}
T.S. Ghiasi, A.A. Kaverzin, P.J. Blah, and
B.J. van Wees, Charge-to-Spin Conversion by the Rashba–Edelstein Effect in Two-Dimensional van der Waals Heterostructures up to Room Temperature, Nano Lett. 19 (2019) 5959.

\bibitem{CISPExpva2}
L. Li, J. Zhang, G. Myeong, W. Shin, H. Lim, B. Kim,
S. Kim, T. Jin, S. Cavill, B.S. Kim, and et al., Gate-Tunable Reversible Rashba-Edelstein Effect in a Few-Layer Graphene/$2$H- TaS$_2$ Heterostructure at Room Temperature, ACS Nano 14 (5) (2020) 5251-5259.

\bibitem{CISPExpHevy1}
I.M. Miron, T. Moore, H. Szambolics, L.D. Buda-
Prejbeanu, S. Auffret, B. Rodmacq, S. Pizzini, J. Vogel,
M. Bonfim, A. Schuhl, and et al., Fast current-induced domain-wall motion controlled by the Rashba effect, Nat. Mater. 10 (6) (2011) 419-423.

\bibitem{CISPExpHevy2}
H.J. Zhang, S. Yamamoto, Y. Fukaya, M. Maekawa,
H. Li, A. Kawasuso, T. Seki, E. Saitoh, and K. Takanashi, Current-induced spin polarization on metal surfaces probed by spin-polarized positron beam, Sci. Rep. 4 (2014) 4844.

\bibitem{CISPExpHevy3}
H.J. Zhang, S. Yamamoto, B. Gu, H. Li, M. Maekawa, Y. Fukaya, and A. Kawasuso, Charge-to-Spin Conversion and Spin Diffusion in Bi/Ag Bilayers Observed by Spin-Polarized Positron Beam, Phys. Rev. Lett. 114 (2025) 166602.

\bibitem{CISPTheo2DElec1}
A. Manchon and S. Zhang, Theory of nonequilibrium intrinsic spin torque in a single nanomagnet, Phys. Rev. B 78 (2008) 212405.

\bibitem{CISPTheo2DElec2}
C. Gorini, A. Maleki Sheikhabadi, K. Shen, I.V. Tokatly, G. Vignale, and R. Raimondi, Theory of current-induced spin polarization in an electron gas, Phys. Rev. B 95 (20) (2017) 205424.

\bibitem{CISPTheo2DElec3}
A. Droghetti and I.V. Tokatly, Current-induced spin polarization at metallic surfaces from first principles, Phys. Rev. B 107 (2023) 174433.

\bibitem{CISPTheo2DElec4}
S. Sun, Y. Ren, D. Zhang, W.-K. Lou, and K. Chang, Current-induced spin polarization and spin Hall effect in III-V compounds two-dimensional materials, Phys. Rev. B 110 (2024) 165430.

\bibitem{CISPTheo3D1}
I.V. Tokatly, E.E. Krasovskii, and G. Vignale, Current-induced spin polarization at the surface of metallic films: A theorem and an ab initio calculation, Phys. Rev. B 91 (3) (2015) 035403.

\bibitem{CISPTheoQuanWell}
A. Maleki Sheikhabadi, I. Miatka1, E. Ya. Sherman, and R. Raimondi1, Theory of the inverse spin galvanic effect in quantum wells, Phys. Rev. B 97 (2018) 235412.

\bibitem{CISPTheo3DHole1}
T. Furukawa, Y. Watanabe, N. Ogasawara, K. Kobayashi, and T. Itou, Current-induced magnetization caused by crystal chirality in nonmagnetic elemental tellurium, Phys. Rev. Res.  3 (2021) 023111.

\bibitem{CISPTheo3DHole2}
K. Tenzin, A. Roy, H. Jafari, B. Banas, F.T. Cerasoli, M. Date, A. Jayaraj, M. Buongiorno Nardelli, and J. Sławinska, Analogs of Rashba-Edelstein effect from density functional theory, Phys. Rev. B 107 (16) (2023) 165140.

\bibitem{CISPTheo3DHole3}
R. Gupta  and A. Droghetti, Current-induced spin polarization in chiral tellurium: A first-principles quantum transport study, Phys. Rev. B 109 (2024) 155141.

\bibitem{dyrdal}
A. Dyrdał, J. Barnaś, and V.K. Dugaev, Current-induced spin polarization in graphene due to Rashba spin-orbit interaction, Phys. Rev. B 89 (7) (2014) 075422.


\bibitem{hosseini}
M.V. Hosseini, The influence of anisotropic Rashba spin–orbit coupling on current-induced spin polarization in graphene, J. Phys.: Condens. Matter. 29 (31) (2017) 315502.

\bibitem{Spintronic1}
I. Žutić, J. Fabian, and S. Das Sarma, Spintronics: Fundamentals and applications, Rev. Mod. Phys. 76 (2004) 323–410.

\bibitem{Spintronic2}
M.I. Dyakonov (Ed.), Spin Physics in Semiconductors, Springer, Berlin/Heidelberg, 2008.

\bibitem{d} 
D. Awschalom and N. Samarth, Trend: Spintronics without magnetism, Physics 2 (2009) 50.

\bibitem{SpintronicsTI}
Y. Fan and K.L. Wang, Spintronics Based on Topological Insulators, SPIN 06 (02) (2016) 1640001.

\bibitem{SpinOrTI1}
B.A. Bernevig, T.L. Hughes, and S.C. Zhang, Quantum spin Hall effect and topological phase transition in HgTe quantum wells, Science 314 (5806) (2006) 1757-1761.

\bibitem{SpinOrTI2}
M. König, S. Wiedmann, C. Brüne, A. Roth, H. Buhmann, L.W. Molenkamp, X.-L. Qi, and S.-C. Zhang, Quantum spin Hall insulator state in HgTe quantum wells, Science 318 (5851) (2007) 766-770.

\bibitem{SwichCISPTI}
F. Yang, S. Ghatak, A.A. Taskin, K. Segawa, Y. Ando, M. Shiraishi, Y. Kanai, K. Matsumoto, A. Rosch, and Y. Ando, Switching of charge-current-induced spin polarization in the topological insulator BiSbTeSe$_2$, Phys. Rev. B 94 (2016) 075304.

\bibitem{OriginCISPTI}
A. Dankert, P. Bhaskar, D. Khokhriakov, I.H. Rodrigues, B. Karpiak, M.V. Kamalakar, S. Charpentier, I. Garate, and S.P. Dash, Origin and evolution of surface spin current in topological insulators, Phys. Rev. B 97 (2018) 125414.

\bibitem{DetectionCISPTI}
C.H. Li, O.M.J. van ‘t Erve, C. Yan, L. Li, and B.T. Jonker, Electrical Detection of Charge-to-spin and Spin-to-Charge Conversion in a Topological Insulator Bi$_2$Te$_3$ Using BN/Al$_2$O$_3$ Hybrid Tunnel Barrier, Sci. Rep. 8  (2018) 10265.

\bibitem{interfaceCISPTI}
C.H. Li, O.M.J. van ‘t Erve, C. Yan, L. Li, and B.T. Jonker, Electrical detection of current generated spin in topological insulator surface states: Role of interface resistance, Sci. Rep. 9 (2019) 6906. 

\bibitem{RoomTemCISPTI1}
Z. Kovács-Krausz, A.M. Hoque, P. Makk, B. Szentpéteri, M. Kocsis, B. Fülöp, M.V. Yakushev, T.V. Kuznetsova, O.E. Tereshchenko, K.A. Kokh, and et al., Electrically Controlled Spin Injection from Giant Rashba Spin–Orbit Conductor BiTeBr, Nano Lett. 20 (7) (2020) 4782–4791.

\bibitem{RoomTemCISPTI2}
M.A. Hoque, L. Sjöström, D. Khokhriakov, B. Zhao, and S.P. Dash, Room temperature nonlocal detection of charge-spin interconversion in a topological insulator, npj 2D Materials and Applications 8 (1) (2024) 10.

\bibitem{TISpinTrans}
L.T. Dang, O. Breunig, Z. Wang, H.F. Legg, and Y. Ando, Topological-Insulator Spin Transistor, Phys. Rev. Appl. 20 (2) (2023) 024065.

\bibitem{CISPPhoton}
J. Yu, H. Zhuang, K. Zhu, Y. Chen, Y. Liu, Y. Zhang, C. Yin, S. Cheng, Y. Lai, K. He, and et al., Observation of current-induced spin polarization in the topological insulator Bi$_2$Te$_3$ via circularly polarized photoconductive differential current, Phys. Rev. B 104 (2021) 045428.

\bibitem{CISPReson}
R. Dey, A. Roy, L.F. Register, and S.K. Banerjee, Recent progress on measurement of spin–charge interconversion in topological insulators using ferromagnetic resonance, APL Mater. 9 (2021) 060702.

\bibitem{CISPThickOpposite}
J. Tian, C. Chang, H. Cao, K. He, X. Ma, Q. Xue, and Y. P. Chen,
Opposite current-induced spin polarization in bulk-insulating and bulk-metallic topological insulators,
Phys. Rev. B 103 (2021) 035412.

\bibitem{comparisonCISP}
C.H. Li, O.M.J. van ‘t Erve, S. Rajput, L. Li, and B.T. Jonker, Direct comparison of current-induced spin polarization in topological insulator Bi$_2$Se$_3$ and InAs Rashba states, Nat. Commun. 7, 13518 (2016).

\bibitem{comparisonCISP0}
Y. Lee, Jonghoon Kim, Seungwon Rho, Seok-Bo Hong, 
Hyeongmun Kim, Jaehan Park, Dajung Kim, Chul Kang, Myung-Ho Bae, Mann-Ho Cho, A comparative study on spin-to-charge and charge-to-spin conversion using modulated Dirac surface states of Bi$_2$Se$_3$, Appl. Surf. Sci. Adv. 25 (2025) 100693.

\bibitem{CISPBiSbTeSe}
C. Hwang, J. Park, S. Ryu, D. H. Shin, and H. C. Koo,
Electrically detected spin polarization in bulk-insulating Bi$_{1.5}$Sb$_{0.5}$Te$_{1.7}$Se$_{1.3}$, Curr. Appl. Phys. 19 (2019) 1155–1159.

\bibitem{CISPBurkov}
A.A. Burkov and D.G. Hawthorn, Spin and Charge Transport on the Surface of a Topological Insulator, Phys. Rev. Lett. 105 (2010) 066802.

\bibitem{CulcerCISP}
D. Culcer, Linear response theory of the Edelstein effect in topological insulator surface states, Physica E 44 (2012) 860.

\bibitem{ConversionCISPTI}
S. Zhang and A. Fert, Conversion between spin and charge currents with topological insulators, Phys. Rev. B 94 (2016) 184423.

\bibitem{interpretCISPTI}
P. Li and I. Appelbaum, Interpreting current-induced spin polarization in topological insulator surface states, Phys. Rev. B 93 (2016) 220404(R).

\bibitem{SemiCISP}
W. Chen, Edelstein and inverse Edelstein effects caused by the pristine surface states of topological insulators, J. Phys. Condens. Matter 32 (2020) 035809.

\bibitem{anisotropicEdelstein}
A.G. Moghaddam, A. Qaiumzadeh, A. Dyrdał, and J. Berakdar, Highly tunable spin-orbit torque and anisotropic magnetoresistance in a topological insulator thin film attached to ferromagnetic layer, Phys. Rev. Lett. 125 (2020) 196801. 

\bibitem{BMR}
A. Dyrdał, J. Barnaś, and A. Fert, Spin-Momentum-Locking Inhomogeneities as a Source of Bilinear Magnetoresistance in Topological Insulators, Phys. Rev. Lett. 124 (2020) 046802.

\bibitem{CISPeffic0}
J.-C. Rojas-Sánchez and A. Fert, Compared Efficiencies of Conversions between Charge and Spin Current by Spin-Orbit Interactions in Two- and Three-Dimensional Systems, 
Phys. Rev. Applied 11 (2019) 054049.

\bibitem{CISPeffic}
A. Johansson, J. Henk, and I. Mertig, Theoretical aspects of the Edelstein effect for anisotropic two-dimensional electron gas and topological insulators, Phys. Rev. B 93 (2016) 195440.

\bibitem{CISPThinFilm}
R. Dey, A. Roy, T. Pramanik, A. Rai, S.H. Shin, S. Majumder, L.F. Register, and S.K. Banerjee, Detection of current induced spin polarization in epitaxial Bi$_2$Te$_3$ thin film, Appl. Phys. Lett. 110 (2017) 122403.

\bibitem{EffecHam1}
R. Yu, W. Zhang, H.-J. Zhang, S.-C. Zhang, X. Dai, and Z. Fang, Quantized anomalous Hall effect in magnetic topological insulators, Science 329 (5987) (2010) 61-64.

\bibitem{EffecHam2}
J. Zhang, C.-Z. Chang, Z. Zhang, J. Wen, X. Feng, K. Li, M. Liu, K. He, L. Wang, X. Chen, and et al., Band structure engineering in (Bi$_1$$_-$$_x$Sb$_x$)$_2$Te$_3$ ternary topological insulators, Nat. Commun. 2 (2011) 574.

\bibitem{highly48}
W.-Y. Shan, H.-Z. Lu, and S.-Q. Shen, Effective continuous model for surface states and thin films of three-dimensional topological insulators, New J. Phys. 12 (4) (2010) 043048. 

\bibitem{thinFiTopo}
H.-Z Lu, W.-Y. Shan, W. Yao, Q. Niu, and S.-Q. Shen, Massive Dirac fermions and spin physics in an ultrathin film of topological insulator,Phys. Rev. B 81 (2010) 115407.

\bibitem{SpiSusc1}
G.D. Mahan, Many-Particle Physics, Kluwer Academic, New York, 2000.

\bibitem{SpiSusc2}
P. Streda, Theory of quantised Hall conductivity in two dimensions, J. Phys. C  15 (1982) L717.

\bibitem{SpiSusc3}
L. Smrcka and P. Streda, Transport coefficients in strong magnetic fields, J. Phys. C 10 (12) (1977) 2153.

\bibitem{SpiSuscVertexCorr}
N.A. Sinitsyn, J.E. Hill, H. Min, J. Sinova,  and A.H. MacDonald, Charge and spin Hall conductivity in metallic graphene, Phys. Rev. Lett. 97 (10) (2006) 106804. 

\bibitem{(Japanese article)}
T. Chiba, S. Takahashi, and G.E.W. Bauer, Magnetic-proximity-induced magnetoresistance on topological insulators, Phys. Rev. B 95 (2017) 094428. 

\bibitem{Zsurka2024}
E. Zsurka, C. Wang, J. Legendre, Da. Di Miceli, L. Serra, D. Gr\"{u}tzmacher, T.L. Schmidt, P. R\"{u}ßmann, and K. Moors, 
Low-energy modeling of three-dimensional topological insulator nanostructures, Phys. Rev. Materials 8 (2024) 084204.

\bibitem{LiNatNano2014}
C.H. Li, O.M.J. van ’t Erve, J.T. Robinson, Y. Liu, L. Li, and B.T. Jonker, Electrical detection of charge-current-induced spin polarization due to spin-momentum locking in Bi$_2$Se$_3$,
Nat. Nanotechnol. 9 (2014) 218–224.

\bibitem{MellnikNature2014}
A.R. Mellnik, J.S. Lee, A. Richardella, J.L. Grab, P.J. Mintun, M.H. Fischer, A. Vaezi, A. Manchon, E.-A. Kim, N. Samarth, and D.C. Ralph,
Spin-transfer torque generated by a topological insulator, Nature 511 (2014) 449–451.

\end{thebibliography}
\end{document}